\documentclass[proceedings]{JHEP}
\usepackage{epsfig}

\newbox\mybox
         
\newcommand\fverb{\setbox\mybox=\hbox\bgroup\verb}
\newcommand\fverbdo{\egroup\medskip\noindent\fbox{\unhbox\mybox}\ }
\newcommand\fverbit{\egroup\item[\fbox{\unhbox\mybox}]}

\font\beeg=cmr17 scaled 1600		
\newcommand\init[1]{\setbox\mybox=\hbox{{\beeg #1}~}%
		   \noindent\global\hangindent=\wd\mybox\global\hangafter-2%
		   \sc\smash{\llap {\lower 13.2pt \box\mybox}}}

\title{\boldmath
Determination of the weak phase $\gamma=\mbox{arg}(V_{ub}^*)$
\unboldmath}

\author{
Matthias Neubert\\
Newman Laboratory of Nuclear Studies\\
Cornell University, Ithaca, NY 14853, USA\\
E-mail: \email{neubert@mail.lns.cornell.edu}}

\conference{Heavy Flavours 8, Southampton, UK, 1999}

\abstract{
Various strategies for extracting or constraining the weak phase 
$\gamma$ with controlled theoretical uncertainties are reviewed. 
Measurements of the rates for the hadronic decays $B^\pm\to\pi K$ 
provide largely model-independent information on $\gamma$. Hadronic 
uncertainties enter only at the level of nonfactorizable 
contributions to the decay amplitudes that are power-suppressed in 
$\Lambda/m_b$ and, simultaneously, violate SU(3) flavor symmetry or 
are doubly Cabibbo suppressed. In addition, these decays have a rich
potential for probing physics beyond the Standard Model. Alternative
proposals for learning $\gamma$ are also discussed briefly.}

\begin{document} 

{\init The main objectives} of the $B$ factories are to explore the 
physics of CP violation, to determine the flavor parameters of the 
electroweak theory, and to probe for signs of physics beyond the 
Standard Model. This will test the Cabibbo--Kobayashi--Maskawa (CKM) 
mechanism, which predicts that all CP violation results from a single 
complex phase in the quark mixing matrix. Facing the announcement of 
evidence for a CP asymmetry in the decays $B\to J/\psi\,K_S$ by 
the CDF Collaboration \cite{CDF}, the confirmation of direct CP 
violation in $K\to\pi\pi$ decays by the KTeV and NA48 groups
\cite{KTeV,NA48}, and the successful start of the asymmetric $B$ 
factories at SLAC and KEK, the year 1999 has been an important step 
in achieving this goal.

The determination of the sides and angles of the ``unitarity 
triangle'' $V_{ub}^* V_{ud}+V_{cb}^* V_{cd}+V_{tb}^* V_{td}=0$ plays 
a central role in the $B$-factory program. Adopting the standard 
phase conventions for the CKM matrix, only the two smallest elements 
in this relation, $V_{ub}^*$ and $V_{td}$, have nonvanishing 
imaginary parts (to an excellent approximation). In the Standard Model 
the angle $\beta=-\mbox{arg}(V_{td})$ can be determined in a 
theoretically clean way by measuring the mixing-induced CP asymmetry 
in the decays $B\to J/\psi\,K_S$. The preliminary CDF result implies 
$\sin2\beta=0.79_{-0.44}^{+0.41}$ \cite{CDF}. The angle 
$\gamma=\mbox{arg}(V_{ub}^*)$, or equivalently the combination 
$\alpha=180^\circ-\beta-\gamma$, is much harder to determine 
\cite{BaBar}. Recently, there has been significant progress in the 
theoretical understanding of the hadronic decays $B\to\pi K$, and 
methods have been developed to extract information on $\gamma$ from 
rate measurements for these processes. Here we discuss the charged 
modes $B^\pm\to\pi K$, which from a theoretical perspective are 
particularly clean. We will, however, also survey alternative methods 
for learning $\gamma$ from other decays.

In the Standard Model the main contributions to the decay amplitudes 
for the rare processes $B\to\pi K$ are due to the penguin-induced 
flavor-changing neutral current (FCNC) transitions 
$\bar b\to\bar s q\bar q$, which exceed a small, Cabibbo-suppressed 
$\bar b\to\bar u u\bar s$ contribution from $W$-boson exchange. The 
weak phase $\gamma$ enters through the interference of these two 
(``penguin'' and ``tree'') contributions. Because of a fortunate 
interplay of isospin, Fierz and flavor symmetries, the theoretical 
description of the charged modes $B^\pm\to\pi K$ is very clean 
despite the fact that these are exclusive nonleptonic decays 
\cite{us,us2,me}. Without any dynamical assumption, the hadronic 
uncertainties in the description of the interference terms relevant 
to the determination of $\gamma$ are of relative magnitude 
$O(\lambda^2)$ or $O(\epsilon_{\rm SU(3)}/N_c)$, where 
$\lambda=\sin\theta_C\approx 0.22$ is a measure of Cabibbo 
suppression, $\epsilon_{\rm SU(3)}\sim 20\%$ is the typical size 
of SU(3) breaking, and the factor $1/N_c$ indicates that the 
corresponding terms vanish in the factorization approximation. 
Factorizable SU(3) breaking can be accounted for in a 
straightforward way. 

Recently, the accuracy of this description has been further 
improved when it was shown that nonleptonic $B$ decays into two light 
mesons, such as $B\to\pi K$ and $B\to\pi\pi$, admit a heavy-quark 
expansion \cite{fact}. To leading order in $\Lambda/m_b$, but to all 
orders in perturbation theory, the decay amplitudes for these 
processes can be calculated from first principles without recourse 
to phenomenological models. The QCD factorization theorem proved in 
\cite{fact} improves upon the phenomenological approach of 
``generalized factorization'' \cite{Stech}, which emerges as the 
leading term in the heavy-quark limit. With the help of this theorem 
the irreducible theoretical uncertainties in the description of the 
$B^\pm\to\pi K$ decay amplitudes can be reduced by an extra factor of 
$O(\Lambda/m_b)$, rendering their analysis essentially model 
independent. As a consequence of this fact, and because they are 
dominated by FCNC transitions, the decays $B^\pm\to\pi K$ offer 
a sensitive probe to physics beyond the Standard Model 
\cite{me,Mati,CDK,anom,troja}, much in the same way as the 
``classical'' FCNC processes $B\to X_s\gamma$ or 
$B\to X_s\,l^+ l^-$. 

In the following three sections we discuss how, in the context of 
the Standard Model, $B^\pm\to\pi K$ rate measurements can be used to 
derive bounds on $\gamma$ and to extract its value with controlled 
theoretical uncertainties. We then explore how these analyses 
could be affected by New Physics. Finally, we briefly summarize
alternative methods to determine $\gamma$ using other decay modes.

\section{\boldmath Theory of $B^\pm\to\pi K$ decays\unboldmath}

The hadronic decays $B\to\pi K$ are mediated by a low-energy
effective weak Hamiltonian, whose operators allow for three distinct 
classes of flavor topologies: QCD penguins, trees, and electroweak 
penguins. In the Standard Model the weak couplings associated with 
these topologies are known. From the measured branching ratios one 
can deduce that QCD penguins dominate the $B\to\pi K$ decay 
amplitudes \cite{Digh}, whereas trees and electroweak penguins are 
subleading and of a similar strength \cite{oldDesh}. The theoretical 
description of the two charged modes $B^\pm\to\pi^\pm K^0$ and 
$B^\pm\to\pi^0 K^\pm$ exploits the fact that the amplitudes for 
these processes differ in a pure isospin amplitude $A_{3/2}$, 
defined as the matrix element of the isovector part of the effective 
Hamiltonian between a $B$ meson and the $\pi K$ isospin eigenstate 
with $I=\frac 32$. In the Standard Model the parameters of this 
amplitude are determined, up to an overall strong phase $\phi$, in 
the limit of SU(3) flavor symmetry \cite{us}. Using the QCD 
factorization theorem proved in \cite{fact}, the SU(3)-breaking 
corrections can be calculated in a model-independent way up to 
nonfactorizable terms that are power-suppressed in $\Lambda/m_b$ and 
vanish in the heavy-quark limit. 

A convenient parameterization of the decay amplitudes 
${\cal A}_{+0}\equiv{\cal A}(B^+\to\pi^+ K^0)$ and
${\cal A}_{0+}\equiv -\sqrt2\,{\cal A}(B^+\to\pi^0 K^+)$ is 
\cite{me}
\begin{eqnarray}\label{ampls}
   {\cal A}_{+0} &=& P\,(1-\varepsilon_a\,e^{i\gamma} e^{i\eta})
    \,, \\
   {\cal A}_{0+} &=& P \Big[ 1 - \varepsilon_a\,e^{i\gamma}
    e^{i\eta} - \varepsilon_{3/2}\,e^{i\phi}
    (e^{i\gamma} - \delta_{\rm EW}) \Big] \,,\nonumber
\end{eqnarray}
where $P$ is the dominant penguin amplitude defined as the sum 
of all terms in the $B^+\to\pi^+ K^0$ amplitude not proportional 
to $e^{i\gamma}$, $\eta$ and $\phi$ are strong phases, and 
$\varepsilon_a$, $\varepsilon_{3/2}$ and $\delta_{\rm EW}$ are 
real hadronic parameters. The weak phase $\gamma$ changes sign 
under a CP transformation, whereas all other parameters stay 
invariant. 

Based on a naive quark-diagram analysis one would not expect the 
$B^+\to\pi^+ K^0$ amplitude to receive a contribution from 
$\bar b\to\bar u u\bar s$ tree topologies; however, such a 
contribution can be induced through final-state rescattering or 
annihilation contributions \cite{Blok,Rob,Ge97,Ne97,Fa97,At97}. 
They are parameterized by $\varepsilon_a=O(\lambda^2)$. In the 
heavy-quark limit this parameter can be calculated and is found to 
be very small, $\varepsilon_a\approx-2\%$ \cite{newfact}. In the 
future, it will be possible to put upper and lower bounds on 
$\varepsilon_a$ by comparing the CP-averaged branching ratios for 
the decays $B^\pm\to\pi^\pm K^0$ and $B^\pm\to K^\pm\bar K^0$ 
\cite{Fa97}. Below we assume $|\varepsilon_a|\le 0.1$; however, our 
results will be almost insensitive to this assumption.

The terms proportional to $\varepsilon_{3/2}$ in (\ref{ampls}) 
parameterize the isospin amplitude $A_{3/2}$. The weak phase 
$e^{i\gamma}$ enters through the tree process 
$\bar b\to\bar u u\bar s$, whereas the quantity $\delta_{\rm EW}$ 
describes the effects of electroweak penguins. The parameter 
$\varepsilon_{3/2}$ measures the relative strength of tree and 
QCD penguin contributions. Information about it can be derived by 
using SU(3) flavor symmetry to relate the tree contribution to 
the isospin amplitude $A_{3/2}$ to the corresponding contribution 
in the decay $B^+\to\pi^+\pi^0$. Since the final state $\pi^+\pi^0$ 
has isospin $I=2$, the amplitude for this process does not receive 
any contribution from QCD penguins. Moreover, electroweak penguins 
in $\bar b\to\bar d q\bar q$ transitions are negligibly small. We 
define a related parameter $\bar\varepsilon_{3/2}$ by writing 
$\varepsilon_{3/2}=\bar\varepsilon_{3/2}
\sqrt{1-2\varepsilon_a\cos\eta\cos\gamma+\varepsilon_a^2}$, so
that the two quantities agree in the limit $\varepsilon_a\to 0$. 
In the SU(3) limit this new parameter can be determined 
experimentally form the relation \cite{us}
\begin{equation}\label{eps}
   \bar\varepsilon_{3/2} = R_1
   \left|\frac{V_{us}}{V_{ud}}\right| \left[
   \frac{2\mbox{B}(B^\pm\to\pi^\pm\pi^0)}
        {\mbox{B}(B^\pm\to\pi^\pm K^0)} \right]^{1/2} .
\end{equation}
SU(3)-breaking corrections are described by the factor 
$R_1=1.22\pm 0.05$, which can be calculated in a model-independent 
way using the QCD factorization theorem for nonleptonic decays 
\cite{newfact}. The quoted error is an estimate of the theoretical 
uncertainty due to corrections of $O(\frac{1}{N_c}\frac{m_s}{m_b})$. 
Using preliminary data reported by the CLEO Collaboration \cite{CLEO} 
to evaluate the ratio of the CP-averaged branching ratios in 
(\ref{eps}) we obtain
\begin{equation}\label{epsval}
   \bar\varepsilon_{3/2} = 0.21\pm 0.06_{\rm exp}
   \pm 0.01_{\rm th} \,.
\end{equation}
With a better measurement of the branching ratios the uncertainty 
in $\bar\varepsilon_{3/2}$ will be reduced significantly.

Finally, the parameter
\begin{eqnarray}\label{delta}
   \delta_{\rm EW} &=& R_2\,
    \left| \frac{V_{cb}^* V_{cs}}{V_{ub}^* V_{us}} \right|\,
    \frac{\alpha}{8\pi}\,\frac{x_t}{\sin^2\!\theta_W}
    \left( 1 + \frac{3\ln x_t}{x_t-1} \right) \nonumber\\
   &=& (0.64\pm 0.09)\times\frac{0.085}{|V_{ub}/V_{cb}|} \,,
\end{eqnarray}
with $x_t=(m_t/m_W)^2$, describes the ratio of electroweak penguin 
and tree contributions to the isospin amplitude $A_{3/2}$. In the 
SU(3) limit it is calculable in terms of Standard Model parameters 
\cite{us,Fl96}. SU(3)-breaking as well as small electromagnetic 
corrections are accounted for by the quantity $R_2=0.92\pm 0.09$ 
\cite{me,newfact}. The error quoted in (\ref{delta}) includes the 
uncertainty in the top-quark mass. 

Important observables in the study of the weak phase $\gamma$ are 
the ratio of the CP-averaged branching ratios in the two 
$B^\pm\to\pi K$ decay modes,
\begin{equation}\label{Rst}
   R_* = \frac{\mbox{B}(B^\pm\to\pi^\pm K^0)}
              {2\mbox{B}(B^\pm\to\pi^0 K^\pm)} 
   = 0.75\pm 0.28 \,,
\end{equation}
and a particular combination of the direct CP asymmetries,
\begin{eqnarray}\label{Atil}
   \widetilde A &=& \frac{A_{\rm CP}(B^\pm\to\pi^0 K^\pm)}{R_*}
    - A_{\rm CP}(B^\pm\to\pi^\pm K^0) \nonumber\\
   &=& -0.52\pm 0.42 \,.
\end{eqnarray}
The experimental values of these quantities are derived using
preliminary CLEO data reported in \cite{CLEO}. The theoretical 
expressions for $R_*$ and $\widetilde A$ obtained using the 
parameterization in (\ref{ampls}) are
\begin{eqnarray}\label{expr}
   R_*^{-1} &=& 1 + 2\bar\varepsilon_{3/2}\cos\phi\,
    (\delta_{\rm EW}-\cos\gamma) \nonumber\\
   &+& \bar\varepsilon_{3/2}^2
    (1-2\delta_{\rm EW}\cos\gamma+\delta_{\rm EW}^2)
    \!+\! O(\bar\varepsilon_{3/2}\varepsilon_a) \,, \nonumber\\
   \widetilde A &=& 2\bar\varepsilon_{3/2} \sin\gamma \sin\phi
    + O(\bar\varepsilon_{3/2}\varepsilon_a) \,.
\end{eqnarray}
Note that the rescattering effects described by $\varepsilon_a$ 
are suppressed by a factor of $\bar\varepsilon_{3/2}$ and thus 
reduced to the percent level. Explicit expressions for these
contributions can be found in \cite{me}.

\section{\boldmath Lower bound on $\gamma$ and constraint in the 
$(\bar\rho,\bar\eta)$ plane\unboldmath}
\label{sec:bound}

There are several strategies for exploiting the above relations. 
From a measurement of the ratio $R_*$ alone a bound on $\cos\gamma$ 
can be derived, implying a nontrivial constraint on the Wolfenstein 
parameters $\bar\rho$ and $\bar\eta$ defining the apex of the 
unitarity triangle \cite{us}. Only CP-averaged branching ratios are 
needed for this purpose. Varying the strong phases $\phi$ and $\eta$ 
independently we first obtain an upper bound on the inverse of 
$R_*$. Keeping terms of linear order in $\varepsilon_a$ yields 
\cite{me}
\begin{eqnarray}\label{Rbound}
   R_*^{-1} &\le& \left( 1 + \bar\varepsilon_{3/2}\,
    |\delta_{\rm EW}-\cos\gamma| \right)^2
    + \bar\varepsilon_{3/2}^2\sin^2\!\gamma \nonumber\\
   &&\mbox{}+ 2\bar\varepsilon_{3/2}|\varepsilon_a|\sin^2\!\gamma \,.
\end{eqnarray}
Provided $R_*$ is significantly smaller than 1, this bound implies 
an exclusion region for $\cos\gamma$ which becomes larger the 
smaller the values of $R_*$ and $\bar\varepsilon_{3/2}$ are. It is 
convenient to consider instead of $R_*$ the related quantity 
\cite{troja}
\begin{equation}\label{Rval}
   X_R = \frac{\sqrt{R_*^{-1}}-1}{\bar\varepsilon_{3/2}} 
   = 0.72\pm 0.98_{\rm exp}\pm 0.03_{\rm th} \,.
\end{equation}
Because of the theoretical factor $R_1$ entering the definition of 
$\bar\varepsilon_{3/2}$ in (\ref{eps}) this is, strictly speaking, 
not an observable. However, the theoretical uncertainty in $X_R$ is 
so much smaller than the present experimental error that it can be
ignored. The advantage of presenting our results in terms of $X_R$ 
rather than $R_*$ is that the leading dependence on 
$\bar\varepsilon_{3/2}$ cancels out, leading to the simple bound 
$|X_R|\le|\delta_{\rm EW}-\cos\gamma|
+O(\bar\varepsilon_{3/2},\varepsilon_a)$.

\FIGURE{
\epsfig{file=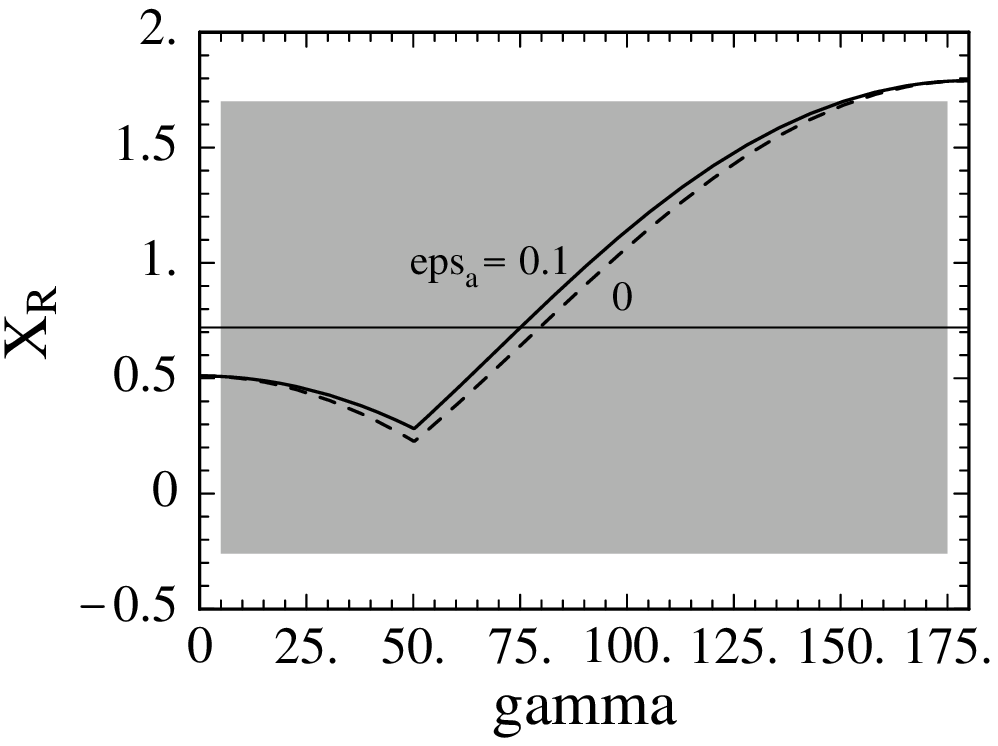,width=6.0cm} 
\caption{
Theoretical upper bound on the ratio $X_R$ versus $|\gamma|$ for 
$\varepsilon_a=0.1$ (solid line) and $\varepsilon_a=0$ (dashed 
line). The horizontal line and band show the current experimental 
value with its $1\sigma$ variation.}
\label{fig:Rbound}}

In Figure~\ref{fig:Rbound} we show the upper bound on $X_R$ as a 
function of $|\gamma|$, obtained by varying the input parameters in 
the intervals $0.15\le\bar\varepsilon_{3/2}\le 0.27$ and 
$0.49\le\delta_{\rm EW}\le 0.79$ (corresponding to using 
$|V_{ub}/V_{cb}|=0.085\pm 0.015$ in (\ref{delta})). Note that the 
effect of the rescattering contribution parameterized by 
$\varepsilon_a$ is very small. The gray band shows the current 
value of $X_R$, which clearly has too large an error to provide any 
useful information on $\gamma$.\footnote{Unfortunately, the 
$2\sigma$ deviation from 1 indicated by a first preliminary CLEO 
result has not been confirmed by the present data.} 
The situation may change, however, once a more precise measurement 
of $X_R$ will become available. For instance, if the current central 
value $X_R=0.72$ were confirmed, it would imply the bound 
$|\gamma|>75^\circ$, marking a significant improvement over the 
indirect limit $|\gamma|>37^\circ$ inferred from the global analysis 
of the unitarity triangle including information from $K$--$\bar K$ 
mixing \cite{BaBar}. 

\FIGURE{
\epsfig{file=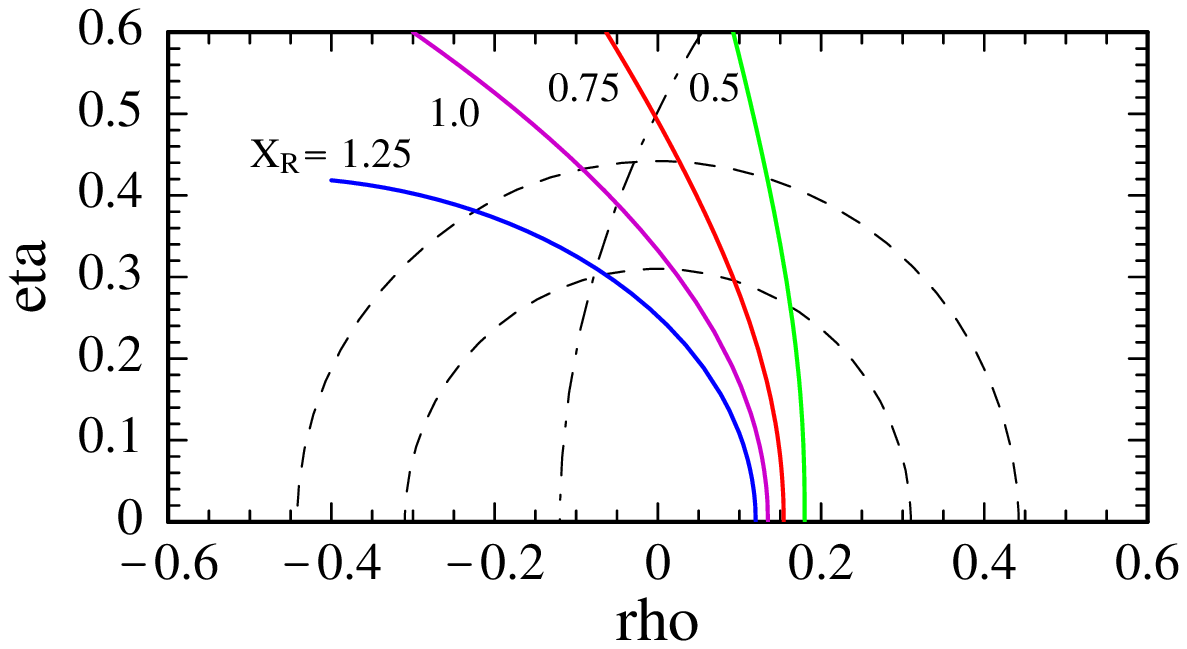,width=6.9cm} 
\caption{
Theoretical constraints on the Wolfenstein parameters 
$(\bar\rho,\bar\eta)$ implied by a measurement of the ratio $X_R$ in 
$B^\pm\to\pi K$ decays (solid lines), semileptonic $B$ decays (dashed 
circles), and $B_{d,s}$--$\bar B_{d,s}$ mixing (dashed-dotted line).}
\label{fig:CKMbound}}

So far, as in previous work, we have used the inequality 
(\ref{Rbound}) to derive a lower bound on $|\gamma|$. However, a 
large part of the uncertainty in the value of $\delta_{\rm EW}$, 
and thus in the resulting bound on $|\gamma|$, comes from the 
present large error on $|V_{ub}|$. Since this is not a hadronic 
uncertainty, it is appropriate to separate it and turn 
(\ref{Rbound}) into a constraint on the Wolfenstein parameters 
$\bar\rho$ and $\bar\eta$. To this end, we use that $\cos\gamma
=\bar\rho/\sqrt{\bar\rho^2+\bar\eta^2}$ by definition, and 
$\delta_{\rm EW}=(0.24\pm 0.03)/\sqrt{\bar\rho^2+\bar\eta^2}$ 
from (\ref{delta}). The solid lines in Figure~\ref{fig:CKMbound} 
show the resulting constraint in the $(\bar\rho,\bar\eta)$ plane 
obtained for the representative values $X_R=0.5$, 0.75, 1.0, 1.25 
(from right to left), which for $\bar\varepsilon_{3/2}=0.21$
would correspond to $R_*=0.82$, 0.75, 0.68, 0.63, respectively. 
Values to the right of these lines are excluded. For comparison, the 
dashed circles show the constraint arising from the measurement of 
the ratio $|V_{ub}/V_{cb}|=0.085\pm 0.015$ in semileptonic $B$ 
decays, and the dashed-dotted line shows the bound implied by the 
present experimental limit on the mass difference $\Delta m_s$ in the 
$B_s$ system \cite{BaBar}. Values to the left of this line are 
excluded. It is evident from the figure that the bound resulting from 
a measurement of the ratio $X_R$ in $B^\pm\to\pi K$ decays may be 
very nontrivial and, in particular, may eliminate the possibility 
that $\gamma=0$. The combination of this bound with information from 
semileptonic decays and $B$--$\bar B$ mixing alone would then 
determine the Wolfenstein parameters $\bar\rho$ and $\bar\eta$ 
within narrow ranges,\footnote{An observation of CP violation, such 
as the measurement of $\epsilon_K$ in $K$--$\bar K$ mixing or 
$\sin2\beta$ in $B\to J/\psi\,K_S$ decays, is however needed to 
fix the sign of $\bar\eta$.}
and in the context of the CKM model would prove the existence of 
direct CP violation in $B$ decays.

\section{\boldmath Extraction of $\gamma$\unboldmath}
\label{sec:determ}

Ultimately, the goal is of course not only to derive a bound on 
$\gamma$ but to determine this parameter directly from the data. 
This requires to fix the strong phase $\phi$ in (\ref{expr}), which 
can be achieved either through the measurement of a CP asymmetry or 
with the help of theory. A strategy for an experimental determination 
of $\gamma$ from $B^\pm\to\pi K$ decays has been suggested in 
\cite{us2}. It generalizes a method proposed by Gronau, Rosner and 
London \cite{GRL} to include the effects of electroweak penguins. The 
approach has later been refined to account for rescattering 
contributions to the $B^\pm\to\pi^\pm K^0$ decay amplitudes 
\cite{me}. Before discussing this method, we will first illustrate an 
easier strategy for a theory-guided determination of $\gamma$ based 
on the QCD factorization theorem for nonleptonic decays \cite{fact}. 
This method does not require any measurement of a CP asymmetry.

\subsection{Theory-guided determination}
\label{sec:thgam}

In the previous section the theoretical predictions for the 
nonleptonic $B\to\pi K$ decay amplitudes obtained using the QCD 
factorization theorem were used in a minimal way, i.e., only to 
calculate the size of the SU(3)-breaking effects parameterized by 
$R_1$ and $R_2$. The resulting bound on $\gamma$ and the 
corresponding constraint in the $(\bar\rho,\bar\eta)$ plane are 
therefore theoretically very clean. However, they are only useful if 
the value of $X_R$ is found to be larger than about 0.5 (see 
Figure~\ref{fig:Rbound}), in which case values of $|\gamma|$ below 
$65^\circ$ are excluded. If it would turn out that $X_R<0.5$, then it 
is possible to satisfy the inequality (\ref{Rbound}) also for small 
values of $\gamma$, however, at the price of having a very large 
strong phase, $\phi\approx 180^\circ$. But this possibility can be 
discarded based on the model-independent prediction that \cite{fact}
\begin{equation}\label{phiest}
   \phi = O[\alpha_s(m_b),\Lambda/m_b] \,.
\end{equation}
A direct calculation of this phase to leading power in $\Lambda/m_b$ 
yields $\phi\approx-11^\circ$ \cite{newfact}. Using the fact that 
$\phi$ is parametrically small, we can exploit a measurement of the 
ratio $X_R$ to obtain a {\em determination\/} of $|\gamma|$ -- 
corresponding to an allowed region in the $(\bar\rho,\bar\eta)$ plane 
-- rather than just a bound. This determination is unique up to a 
sign. Note that for small values of $\phi$ the impact of the strong 
phase in the expression for $R_*$ in (\ref{expr}) is a second-order 
effect. As long as $|\phi|\ll\sqrt{2\Delta\bar\varepsilon_{3/2}/
\bar\varepsilon_{3/2}}$, the uncertainty in $\cos\phi$ has a much 
smaller effect than the uncertainty in $\bar\varepsilon_{3/2}$. With 
the present value of $\bar\varepsilon_{3/2}$ this is the case as long 
as $|\phi|\ll 43^\circ$. We believe it is a safe assumption to take 
$|\phi|<25^\circ$ (i.e., more than twice the value obtained to 
leading order in $\Lambda/m_b$), so that $\cos\phi>0.9$. 

\FIGURE{
\epsfig{file=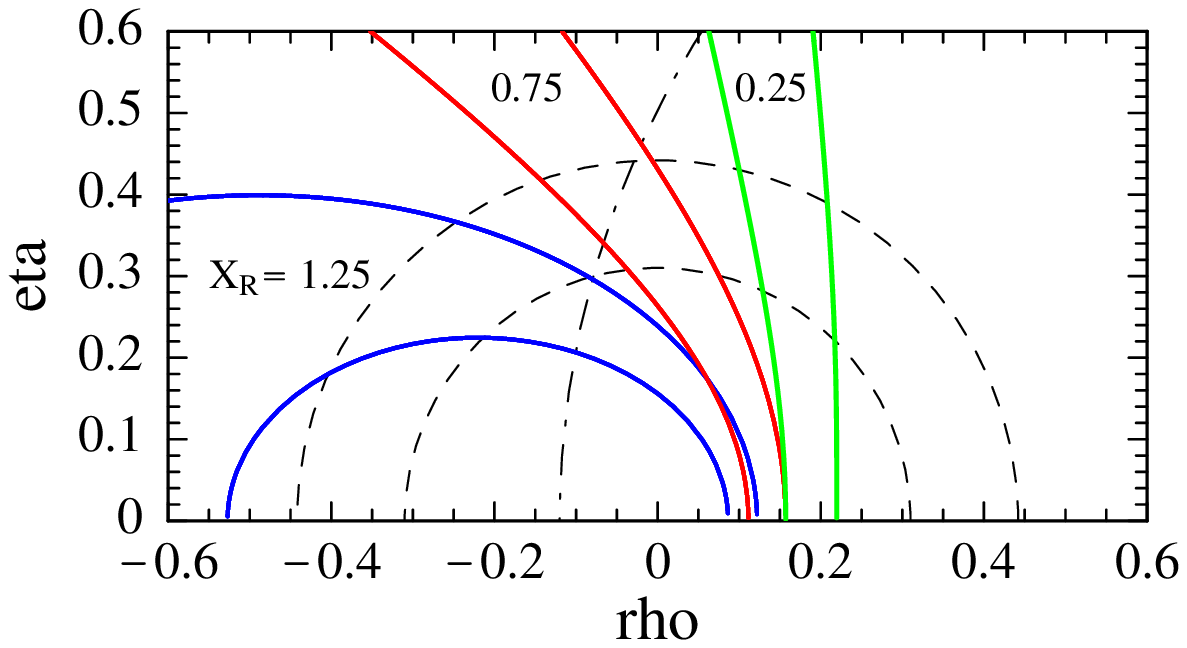,width=6.9cm} 
\caption{
Allowed regions in the $(\bar\rho,\bar\eta)$ plane for fixed values 
of $X_R$, obtained by varying all theoretical parameters inside 
their respective ranges of uncertainty, as specified in the text. 
The sign of $\bar\eta$ is not determined.}
\label{fig:CKMfit}}

Solving the equation for $R_*$ in (\ref{expr}) for $\cos\gamma$, and 
including the corrections of $O(\varepsilon_a)$, we find 
\begin{eqnarray}\label{gamth}
   \cos\gamma &=& \delta_{\rm EW}
    - \frac{X_R + \frac12\bar\varepsilon_{3/2}
            (X_R^2-1+\delta_{\rm EW}^2)}
           {\cos\phi+\bar\varepsilon_{3/2}\delta_{\rm EW}}
    \nonumber\\
   &&\mbox{}+ 
    \frac{\varepsilon_a\cos\eta\sin^2\!\gamma}
         {\cos\phi+\bar\varepsilon_{3/2}\delta_{\rm EW}} \,,
\end{eqnarray}
where we have set $\cos\phi=1$ in the numerator of the 
$O(\varepsilon_a)$ term. Using the QCD factorization theorem one 
finds that $\varepsilon_a\cos\eta\approx -0.02$ in the heavy-quark 
limit \cite{newfact}, and we assign a 100\% uncertainty to this 
estimate. In evaluating the result (\ref{gamth}) we scan the 
parameters in the ranges $0.15\le\bar\varepsilon_{3/2}\le 0.27$, 
$0.55\le\delta_{\rm EW}\le 0.73$, $-25^\circ\le\phi\le 25^\circ$, 
and $-0.04\le\varepsilon_a\cos\eta\sin^2\!\gamma\le 0$. 
Figure~\ref{fig:CKMfit} shows the allowed regions in the 
$(\bar\rho,\bar\eta)$ plane for the representative values $X_R=0.25$, 
0.75, and 1.25 (from right to left). We stress that with this method 
a useful constraint on the Wolfenstein parameters is obtained for 
{\em any\/} value of $X_R$.

\subsection{Model-independent determination}

\FIGURE{
\epsfig{file=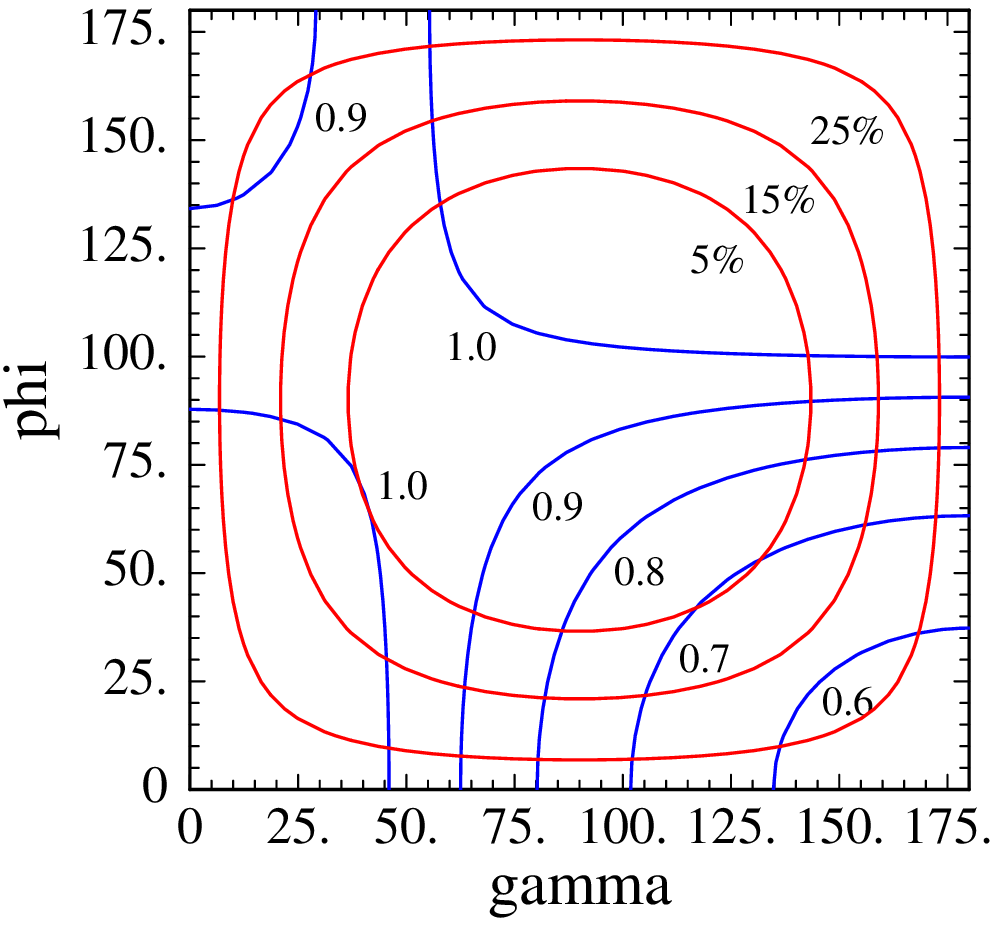,width=6.0cm} 
\caption{
Contours of constant $R_*$ (``hyperbolas'') and constant 
$|\widetilde A|$ (``circles'') in the $(|\gamma|,|\phi|)$ plane. 
The sign of the asymmetry $\widetilde A$ determines the sign of the 
product $\sin\gamma\sin\phi$. The contours for $R_*$ refer to values 
from 0.6 to 1.0 in steps of 0.1, those for the asymmetry correspond 
to 5\%, 15\%, and 25\%, as indicated.}
\label{fig:contours}}

It is important that, once more precise data on $B^\pm\to\pi K$ 
decays will become available, it will be possible to test the
prediction of a small strong phase $\phi$ experimentally. To this 
end, one must determine the CP asymmetry $\widetilde A$ defined in 
(\ref{Atil}) in addition to the ratio $R_*$. From (\ref{expr}) it 
follows that for fixed values of $\bar\varepsilon_{3/2}$ and 
$\delta_{\rm EW}$ the quantities $R_*$ and $\widetilde A$ define 
contours in the $(\gamma,\phi)$ plane, whose intersections determine 
the two phases up to possible discrete ambiguities \cite{us2,me}. 
Figure~\ref{fig:contours} shows these contours for some 
representative values, assuming $\bar\varepsilon_{3/2}=0.21$, 
$\delta_{\rm EW}=0.64$, and $\varepsilon_a=0$. In practice, including 
the uncertainties in the values of these parameters changes the 
contour lines into contour bands. Typically, the spread of the bands 
induces an error in the determination of $\gamma$ of about $10^\circ$ 
\cite{me}.\footnote{A precise determination of this error requires 
know\-ledge of the actual values of the observables. Gronau and Pirjol 
\protect\cite{Pirj} find a larger error for the special case where 
the product $|\sin\gamma\sin\phi|$ is very close to 1, which however 
is highly disfavored because of the expected smallness of the strong 
phase $\phi$.} 
In the most general case there are up to eight discrete solutions for 
the two phases, four of which are related to the other four by a sign 
change $(\gamma,\phi)\to(-\gamma,-\phi)$. However, for typical values 
of $R_*$ it turns out that often only four solutions exist, two of 
which are related to the other two by a sign change. The theoretical 
prediction that $\phi$ is small implies that solutions should exist 
where the contours intersect close to the lower portion in the plot. 
Other solutions with large $\phi$ are strongly disfavored. Note that 
according to (\ref{expr}) the sign of the CP asymmetry $\widetilde A$ 
fixes the relative sign between the two phases $\gamma$ and $\phi$. 
If we trust the theoretical prediction that $\phi$ is negative 
\cite{newfact}, it follows that in most cases there remains only a 
unique solution for $\gamma$, i.e., {\em the CP-violating phase 
$\gamma$ can be determined without any discrete ambiguity}. 

Consider, as an example, the hypothetical case where $R_*=0.8$ and 
$\widetilde A=-15\%$. Figure~\ref{fig:contours} then allows the four 
solutions where $(\gamma,\phi)\approx(\pm 82^\circ,\mp 21^\circ)$ or 
$(\pm 158^\circ,\mp 78^\circ)$. The second pair of solutions is 
strongly disfavored because of the large values of the strong phase 
$\phi$. From the first pair of solutions, the one with 
$\phi\approx-21^\circ$ is closest to our theoretical expectation that 
$\phi\approx -11^\circ$, hence leaving $\gamma\approx 82^\circ$ as 
the unique solution.

\section{Sensitivity to New Physics}
\label{sec:NP}

In the presence of New Physics the theoretical description of 
$B^\pm\to\pi K$ decays becomes more complicated. In particular, new 
CP-violating contributions to the decay amplitudes may be induced. 
A detailed analysis of such effects has been presented in 
\cite{troja}. A convenient and completely general parameterization of 
the two amplitudes in (\ref{ampls}) is obtained by replacing
\begin{eqnarray}\label{replace}
   P &\to& P' \,, \qquad
    \varepsilon_a\,e^{i\gamma} e^{i\eta} \to
    i\rho\,e^{i\phi_\rho} \,, \nonumber\\
   \delta_{\rm EW} &\to& a\,e^{i\phi_a} + ib\,e^{i\phi_b} \,,
\end{eqnarray}
where $\rho$, $a$, $b$ are real hadronic parameters, and $\phi_\rho$, 
$\phi_a$, $\phi_b$ are strong phases. The terms $i\rho$ and $ib$ 
change sign under a CP transformation. New Physics effects 
parameterized by $P'$ and $\rho$ are isospin conserving, while those 
described by $a$ and $b$ violate isospin symmetry. Note that the 
parameter $P'$ cancels in all ratios of branching ratios and thus does 
not affect the quantities $R_*$ and $X_R$ as well as any CP 
asymmetry. Because the ratio $R_*$ in (\ref{Rst}) would be 1 in the 
limit of isospin symmetry, it is particularly sensitive to 
isospin-violating New Physics contributions. 

New Physics can affect the bound on $\gamma$ derived from 
(\ref{Rbound}) as well as the extraction of $\gamma$ using the 
strategies discussed above. We will discuss these two possibilities 
in turn.

\subsection{\boldmath Effects on the bound on $\gamma$\unboldmath}

The upper bound on $R_*^{-1}$ in (\ref{Rbound}) and the 
corresponding bound on $X_R$ shown in Figure~\ref{fig:Rbound} are 
model-independent results valid in the Standard Model. Note that 
the extremal value of $R_*^{-1}$ is such that 
$|X_R|\le(1+\delta_{\rm EW})$ irrespective of $\gamma$. A value of 
$|X_R|$ exceeding this bound would be a clear signal for New Physics 
\cite{me,Mati,troja}. 

Consider first the case where New Physics may induce arbitrary 
CP-violating contributions to the $B\to\pi K$ decay amplitudes, 
while preserving isospin symmetry. Then the only change with respect 
to the Standard Model is that the parameter $\rho$ may no longer be 
as small as $O(\varepsilon_a)$. Varying the strong phases $\phi$ and 
$\phi_\rho$ independently, and allowing for an {\em arbitrarily 
large\/} New Physics contribution to $\rho$, one can derive the 
bound \cite{troja} 
\begin{equation}\label{phiarb}
   |X_R| \le \sqrt{1 - 2\delta_{\rm EW}\cos\gamma 
   + \delta_{\rm EW}^2} \le 1+\delta_{\rm EW} \,.
\end{equation}
Note that the extremal value is the same as in the Standard Model, 
i.e., isospin-conserving New Physics effects cannot lead to a value 
of $|X_R|$ exceeding $(1+\delta_{\rm EW})$. For intermediate values 
of $\gamma$ the Standard Model bound on $X_R$ is weakened; but even 
for large values $\rho=O(1)$, corresponding to a significant New 
Physics contribution to the decay amplitudes, the effect is small.

If both isospin-violating and isospin-conser\-ving New Physics 
contributions are present and involve new CP-violating phases, the 
analysis becomes more complicated. Still, it is possible to derive 
model-independent bounds on $X_R$. Allowing for arbitrary values 
of $\rho$ and all strong phases, one obtains \cite{troja}
\begin{eqnarray}\label{abbound}
   |X_R| &\le& \sqrt{(|a|+|\cos\gamma|)^2 + (|b|+|\sin\gamma|)^2}
    \nonumber\\
   &\le& 1 + \sqrt{a^2 + b^2}
    \le \frac{2}{\bar\varepsilon_{3/2}} + X_R \,,
\end{eqnarray}
where the last inequality is relevant only in cases where
$\sqrt{a^2 + b^2}\gg 1$. The important point to note is that with
isospin-violating New Physics contributions the value of $|X_R|$
can exceed the upper bound in the Standard Model by a potentially 
large amount. For instance, if $\sqrt{a^2+b^2}$ is twice as large
as in the Standard Model, corresponding to a New Physics contribution 
to the decay amplitudes of only 10--15\%, then $|X_R|$ could be as 
large as 2.6 as compared with the maximal value 1.8 allowed in the 
Standard Model. Also, in the most general case where $b$ and $\rho$ 
are nonzero, the maximal value $|X_R|$ can take is no longer 
restricted to occur at the endpoints $\gamma=0^\circ$ or $180^\circ$, 
which are disfavored by the global analysis of the unitarity triangle 
\cite{BaBar}. Rather, $|X_R|$ would take its maximal value if 
$|\tan\gamma|=|\rho|=|b/a|$.

The present experimental value of $X_R$ in (\ref{Rval}) has too 
large an error to determine whether there is any deviation from the 
Standard Model. If $X_R$ turns out to be larger than 1 (i.e., at
least one third of a standard deviation above its current central 
value), then an interpretation of this result in the Standard Model 
would require a large value $|\gamma|>91^\circ$ (see 
Figure~\ref{fig:Rbound}), which would be difficult to accommodate in 
view of the upper bound implied by the experimental constraint on 
$B_s$--$\bar B_s$ mixing, thus providing evidence for New Physics. 
If $X_R>1.3$, one could go a step further and conclude that the New 
Physics must necessarily violate isospin \cite{troja}. 

\subsection{\boldmath Effects on the determination of $\gamma$
\unboldmath}

A value of the observable $R_*$ violating the bound (\ref{Rbound}) 
would be an exciting hint for New Physics. However, even if a future 
precise measurement will give a value that is consistent with the 
Standard Model bound, $B^\pm\to\pi K$ decays provide an excellent 
testing ground for physics beyond the Standard Model. This is so 
because New Physics may cause a significant shift in the value of 
$\gamma$ extracted using the strategies discussed in 
Section~\ref{sec:determ}, leading to inconsistencies when this value 
is compared with other determinations of $\gamma$. 

A global fit of the unitarity triangle combining information from 
semileptonic $B$ decays, $B$--$\bar B$ mixing, CP violation in the 
kaon system, and mixing-induced CP violation in $B\to J/\psi\,K_S$ 
decays provides information on $\gamma$ which in a few years will 
determine its value within a rather narrow range \cite{BaBar}. Such 
an indirect determination could be complemented by direct 
measurements of $\gamma$ using, e.g., $B\to D K^{(*)}$ decays (see 
below), or using the triangle relation $\gamma=180^\circ-\alpha-\beta$ 
combined with a measurement of $\alpha$. We will assume that a 
discrepancy of more than $25^\circ$ between the ``true'' 
$\gamma=\mbox{arg}(V_{ub}^*)$ and the value $\gamma_{\pi K}$ 
extracted in $B^\pm\to\pi K$ decays will be observable after a few 
years of operation at the $B$ factories. This sets the benchmark for 
sensitivity to New Physics effects.

\FIGURE{
\epsfig{file=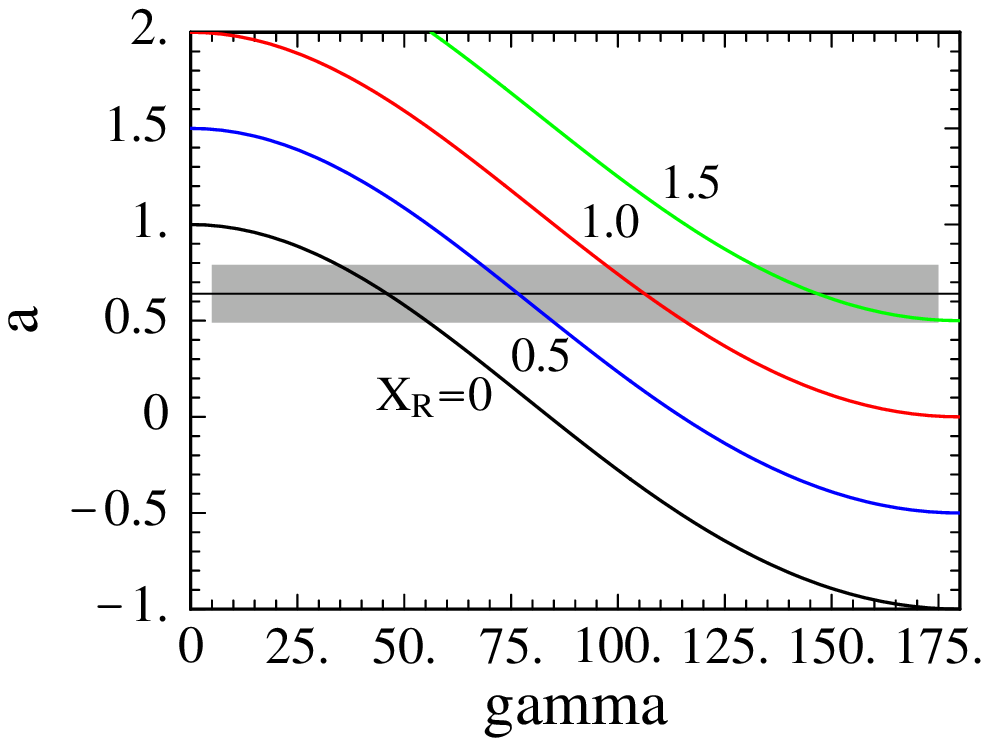,width=6.0cm} 
\caption{
Contours of constant $X_R$ versus $\gamma$ and the parameter $a$, 
assuming $\gamma>0$. The horizontal band shows the value of $a$ 
in the Standard Model.}
\label{fig:shift}}

In order to illustrate how big an effect New Physics could have 
on the extracted value of $\gamma$ we consider the simplest case 
where there are no new CP-violating couplings. Then all New Physics 
contributions in (\ref{replace}) are parameterized by the single 
parameter $a_{\rm NP}\equiv a-\delta_{\rm EW}$. A more general 
discussion can be found in \cite{troja}. We also assume for 
simplicity that the strong phase $\phi$ is small, as suggested by 
(\ref{phiest}). In this case the difference between the value 
$\gamma_{\pi K}$ extracted from $B^\pm\to\pi K$ decays and the 
``true'' value of $\gamma$ is to a good approximation given by
\begin{equation}
   \cos\gamma_{\pi K} \simeq \cos\gamma - a_{\rm NP} \,.
\end{equation}
In Figure~\ref{fig:shift} we show contours of constant $X_R$ versus 
$\gamma$ and $a$, assuming without loss of generality that 
$\gamma>0$. Obviously, even a moderate New Physics contribution to 
the parameter $a$ can induce a large shift in $\gamma$. Note that 
the present central value of $X_R\approx 0.7$ is such that values of 
$a$ less than the Standard Model result $a\approx 0.64$ are 
disfavored, since they would require values of $\gamma$ exceeding 
$100^\circ$, in conflict with the global analysis of the unitarity 
triangle \cite{BaBar}.

\subsection{Survey of New Physics models}

In \cite{troja}, we have explored how physics beyond the Standard 
Model could affect purely hadronic FCNC transitions of the type 
$\bar b\to\bar s q\bar q$ focusing, in particular, on isospin 
violation. Unlike in the Standard Model, where isospin-violating 
effects in these processes are suppressed by electroweak gauge 
couplings or small CKM matrix elements, in many New Physics 
scenarios these effects are not parametrically suppressed relative 
to isospin-conserving FCNC processes. In the language of effective 
weak Hamiltonians this implies that the Wilson coefficients of QCD 
and electroweak penguin operators are of a similar magnitude. For a 
large class of New Physics models we found that the coefficients of 
the electroweak penguin operators are, in fact, due to ``trojan'' 
penguins, which are neither related to penguin diagrams nor of 
electroweak origin. 

Specifically, we have considered: (a) models with tree-level FCNC 
couplings of the $Z$ boson, extended gauge models with an extra $Z'$ 
boson, supersymmetric models with broken R-parity; (b) supersymmetric 
models with R-parity conservation; (c) two-Higgs--doublet models, and 
models with anomalous gauge-boson couplings. Some of these models 
have also been investigated in \cite{CDK,anom}. In case (a), the 
resulting electroweak penguin coefficients can be much larger than in 
the Standard Model because they are due to tree-level processes. In 
case (b), these coefficients can compete with the ones of the 
Standard Model because they arise from strong-interaction box 
diagrams, which scale relative to the Standard Model like
$(\alpha_s/\alpha)(m_W^2/m_{\rm SUSY}^2)$. In models (c), on 
the other hand, isospin-violating New Physics effects are not
parametrically enhanced and are generally smaller than in the 
Standard Model.

For each New Physics model we have explored which region of 
parameter space can be probed by the $B^\pm\to\pi K$ observables, 
and how big a departure from the Standard Model predictions one 
can expect under realistic circumstances, taking into account all 
constraints on the model parameters implied by other processes. 
Table~\ref{tab:1} summarizes our estimates of the maximal 
isospin-violating contributions to the decay amplitudes, as 
parameterized by $|a_{\rm NP}|$. They are the potentially most 
important source of New Physics effects in $B^\pm\to\pi K$ 
decays. For comparison, we recall that in the Standard Model 
$a\approx 0.64$. Also shown are the corresponding maximal values 
of the difference $|\gamma_{\pi K}-\gamma|$. As noted above, in 
models with tree-level FCNC couplings New Physics effects can be 
dramatic, whereas in supersymmetric models with R-parity 
conservation isospin-violating loop effects can be competitive with 
the Standard Model. In the case of supersymmetric models with 
R-parity violation the bound (\ref{abbound}) implies interesting 
limits on certain combinations of the trilinear couplings 
$\lambda_{ijk}'$ and $\lambda_{ijk}''$, as discussed in \cite{troja}.

\TABULAR[t]{|l|cc|}
{\hline
Model & $\!|a_{\rm NP}|\!$ & $\!|\gamma_{\pi K}-\gamma|$ \\
\hline
FCNC $Z$ exchange & 2.0 & $180^\circ$ \\
extra $Z'$ boson & 14$^*$ & $180^\circ$ \\
SUSY without R-parity & 14$^*$ & $180^\circ$ \\
\hline
SUSY with R-parity & 0.4 & $25^\circ$ \\
                   & 1.3 & $180^\circ$ \\
\hline
two-Higgs--doublet mod.\ & 0.15 & $10^\circ$ \\
anom.\ gauge-boson coupl.\ \hspace{-0.2cm} & 0.3 & $20^\circ$ \\
\hline}
{\label{tab:1}
Maximal contributions to $a_{\rm NP}$ and shifts in $\gamma$ in 
extensions of the Standard Model. Entries marked with ``$^*$'' 
are upper bounds derived using (\protect\ref{abbound}). For the 
case of supersymmetric models with R-parity the first (second) 
row corresponds to maximal right-handed (left-handed) 
strange--bottom squark mixing. For the two-Higgs--doublet models
we take $m_{H^+}>100$\,GeV and $\tan\beta>1$.}

\section{Alternative approaches and recent developments}

We will now review recent developments regarding other approaches 
to determining $\gamma$, focusing mainly on proposals that can be 
pursued at the first-generation $B$ factories.

\subsection{\boldmath Variants of the $B^\pm\to\pi K$ strategy
\unboldmath}

The first proposal to constraining $\gamma$ using CP-averaged
$B\to\pi K$ branching ratios was made by Fleischer and Mannel 
\cite{FM}, who suggested to consider the ratio
\begin{equation}
   R = \frac{\tau(B^+)}{\tau(B^0)}\,
   \frac{\mbox{B}(B^0\to\pi^\mp K^\pm)}
        {\mbox{B}(B^\pm\to\pi^\pm K^0)} 
   = 1.11\pm 0.33 \,.
\end{equation}
Neglecting the small rescattering contribution to the 
$B^\pm\to\pi^\pm K^0$ decay amplitudes as well as electroweak 
penguin contributions yields
\begin{eqnarray}\label{FMbound}
   R &\simeq& 1 - 2\varepsilon_T\cos\phi_T\cos\gamma
    + \varepsilon_T^2 \nonumber\\
   &=& \sin^2\!\gamma + (\varepsilon_T-\cos\phi_T\cos\gamma)^2
    + \sin^2\!\phi_T\cos^2\!\gamma \nonumber\\
   &\ge& \sin^2\!\gamma \,,
\end{eqnarray}
where $\varepsilon_T$ is a real parameter of similar magnitude as
$\bar\varepsilon_{3/2}$, and $\phi_T$ is a strong phase.
If the ratio $R$ was found significantly less than 1, the above
inequality would imply an exclusion region around $\gamma=90^\circ$.  

Unlike the parameter $\bar\varepsilon_{3/2}$ in $B^\pm\to\pi K$ 
decays, the quantity $\varepsilon_T$ is not constrained by SU(3) 
symmetry and cannot be determined experimentally. The strategy 
proposed in \cite{FM} is to eliminate this quantity in deriving 
a bound on $\gamma$. This weakens the handle on the weak phase 
except for the particular case where 
$\varepsilon_T\approx\cos\phi_T\cos\gamma$. The neglect of 
electroweak penguin and rescattering corrections is questionable 
and has given rise to some criticism \cite{Ge97,Ne97,Fa97,At97}. 
Yet, although the bound (\ref{FMbound}) is theoretically not as 
clean as the corresponding bound (\ref{Rbound}) on the ratio $R_*$, 
a precise measurement of the ratio $R$ can provide for an 
interesting consistency check. Various refinements and extensions 
of the original Fleischer--Mannel strategy are discussed in 
\cite{Robert}.

Some authors have suggested to eliminate the small rescattering
contribution to the $B^\pm\to\pi^\pm K^0$ decay amplitudes,  
parameterized by $\varepsilon_a$ in (\ref{ampls}), by assuming SU(3) 
symmetry and exploiting amplitude relations connecting $B\to\pi K$ 
and $B\to\pi\pi$ decays with other decay modes, such as 
$B^+\to K^+\bar K^0$, $B^+\to\pi^+\eta^{(\prime)}$, $B_s\to K\pi$ 
\cite{GroPir}, or $B^+\to K^+\eta^{(\prime)}$, 
$B_s\to\pi^0\eta^{(\prime)}$ \cite{AgaDes}. Note that approaches 
based on $B_s$ decays have to await second-generation $B$ factories 
at hadron colliders. Besides relying on the assumption of SU(3) 
flavor symmetry the above proposals suffer from theoretical 
uncertainties related to $\eta$--$\eta'$ mixing. Given that the 
rescattering contribution in question is expected to be very small 
\cite{newfact}, and that this expectation can be tested 
experimentally using $B^\pm\to K^\pm\bar K^0$ decays \cite{Fa97}, 
neglecting $\varepsilon_a$ or putting an upper bound on it will most 
likely be a better approximation than neglecting the potentially much 
larger SU(3)-breaking corrections in the above strategies.

\subsection{\boldmath SU(3) relations between $B_d$ and $B_s$ decay 
amplitudes\unboldmath}

Fleischer has recently suggested to use the U-spin 
($d\leftrightarrow s$) subgroup of flavor SU(3) to derive relations 
between various decays into CP eigenstates, such as \cite{newRob}
\begin{eqnarray}
   B_d\to J/\psi\,K_S &~\leftrightarrow~& B_s\to J/\psi\,K_S \,,
    \nonumber\\
   B_d\to D_d^+ D_d^- &~\leftrightarrow~& B_s\to D_s^+ D_s^- \,,
    \nonumber\\
   B_d\to\pi^+\pi^- &~\leftrightarrow~& B_s\to K^+ K^- \,.
\end{eqnarray}
The first two relations provide access to the weak phase $\gamma$, 
while the third one is sensitive to both $\beta$ and $\gamma$. 
Although this strategy involves $B_s$ decays and thus cannot be 
pursued at the first-generation $B$ factories, we discuss it because 
of its general nature.

Consider the example of $B_d, B_s\to J/\psi\,K_S$ decays, which are 
governed by an interference of tree and penguin topologies. The 
sensitivity to $\gamma$ arises from the presence of top- and up-quark 
penguins. A general parameterization of the decay amplitudes 
$A_{d,s}\equiv{\cal A}(B_{d,s}\to J/\psi\,K_S)$ is
\begin{eqnarray}
   A_d &=& A' \big( 1 + \tan^2\!\theta_C\,\epsilon'\,
    e^{i\phi'} e^{i\gamma} \big) \,, \nonumber\\
   A_s &=& A \left( 1 + \epsilon\,e^{i\phi}\,e^{i\gamma}
    \right) \,,
\end{eqnarray}
where $\theta_C$ is the Cabibbo angle, $\phi^{(\prime)}$ are strong
phases, and the penguin contributions are proportional to parameters 
$\epsilon^{(\prime)}\sim 0.2$. Exact U-spin symmetry would imply 
$A'=A$, $\epsilon'=\epsilon$, and $\phi'=\phi$. In that limit 
$\gamma$ could be determined (with discrete ambiguities) by 
measuring the direct and mixing-induced CP asymmetries 
$A_{\rm CP}^{\rm dir}(B_s\to J/\psi\,K_S)$ and
$A_{\rm CP}^{\rm mix}(B_s\to J/\psi\,K_S)$ as well as the ratio of 
the CP-averaged $B_d, B_s\to J/\psi\,K_S$ decay rates. These
observables would suffice to fix $\epsilon$, $\phi$ and $\gamma$. 
Assuming $A'=A$, this parameter cancels out.

This approach is interesting in that it is general and can be 
applied to several different decay modes \cite{newRob}. Assuming 
the theoretical uncertainties related to U-spin breaking can be 
controlled, it will provide several independent determinations of 
$\beta$ and $\gamma$. However, a question that needs to be addressed 
in future work is how important the SU(3)-breaking corrections 
leading to $A'\ne A$ are, given that $\epsilon\sim 0.2$ is expected 
to be a small parameter. 

\subsection{\boldmath Dalitz-plot analysis in 
$B^\pm\to\pi^\pm\pi^+\pi^-$ decays\unboldmath}

There have been several proposals for obtaining information on the 
weak phases $\alpha$ and $\gamma$ from an analysis of the Dalitz 
plot in $B\to 3\pi$ decays. In fact, the approach of Quinn and Snyder 
\cite{Dalitz} based on the decays $B\to\rho\pi\to 3\pi$ is considered 
to offer one of the most promising ways to measure $\alpha$ 
\cite{Isihere}. The strategy of Bediaga et al.\ \cite{Bedi} (see also 
\cite{Bajc}) for learning $\gamma$ is to fit the measured Dalitz 
distribution $\mbox{d}^2\Gamma/\mbox{d}m_1^2\,\mbox{d}m_2^2$ in the 
decays $B^+\to\pi^+\pi^+\pi^-$, where $m_i^2=(p_{\pi_i^+}
+p_{\pi^-})^2$, to an ansatz of the form
\begin{equation}
   \Big| \sum_i a_i\,e^{i\theta_i}\,F_i(m_1^2,m_2^2) \Big|^2 \,.
\end{equation}
$F_i(m_1^2,m_2^2)$ are appropriate kinematic functions for resonance 
or continuum contributions. From the fit one extracts a set 
$\{a_i,\theta_i\}$ of real amplitudes and complex phases. Performing 
a similar fit to the CP-conjugated decays $B^-\to\pi^-\pi^-\pi^+$ 
gives parameters $\{a_i,\bar\theta_i\}$. The complex phases are sums 
of strong and weak phases, and the latter ones are determined from 
the differences $\phi_i=\frac12(\theta_i-\bar\theta_i)$. The weak 
phase $\gamma$ enters through the interference of the $\chi_c\pi^\pm$ 
resonance state with other resonance channels (e.g., $\rho^0\pi^\pm$, 
$f_0\pi^\pm$, etc.) and nonresonant contributions. The associated 
CKM factors are $V_{cb}^* V_{cd}$ and $V_{ub}^* V_{ud}$, 
respectively.

A theoretical problem inherent in this approach is the ``penguin
pollution'', i.e., the fact that $\bar b\to\bar d q\bar q$ penguin
transitions contaminate the analysis. If the penguin/tree ratio is 
assumed to be at most 20\%, then the resulting error in the 
extraction of $\gamma$ is bound to be less than $11^\circ$ 
\cite{Bedi}. Unfortunately, the recent data on $B\to\pi K$ and 
$B\to\pi\pi$ decays reported by the CLEO Collaboration \cite{CLEO} 
suggest that the penguin/tree ratio may be significantly larger than 
20\%.

The feasibility of this method profits from the fact that no 
tagging is required (only charged $B$-meson decays are considered), 
the final state consists of three charged pions (no $\pi^0$ 
reconstruction is needed), and a Dalitz plot analysis typically does 
not require very large data samples. The authors of \cite{Bedi} 
estimate that with only 1000 events one could obtain a resolution of 
$\Delta\gamma\approx 20^\circ$. Potential problems of the approach 
are that the size of the interference term depends on the yet 
unknown $B^\pm\to\chi_c\pi^\pm\to\pi^\pm\pi^+\pi^-$ branching ratio, 
and that contamination from nonresonant channels may, in the end, 
require larger data samples.

\subsection{\boldmath Extracting $\gamma$ in $B\to D K^{(*)}$ 
decays\unboldmath}

$B\to D K^{(*)}$ decays were originally considered to be the 
``classical'' way for determining $\gamma$. Later, it was realized 
that this is a very challenging route, which poses high demands to 
experiment and theory. We discuss this strategy because it provides 
an extraction of $\gamma$ from tree processes alone, which is 
unlikely to be affected much by New Physics.  

The original idea of Gronau and Wyler \cite{GroW} (see also 
\cite{Isi,AEGS}) was to use the interference of the amplitudes for 
the decays $B^+\to\bar D^0 K^+$ and $B^+\to D^0 K^+$ occurring if the
charm meson is detected as a CP eigenstate $D_1^0=\frac{1}{\sqrt2}
(D^0+\bar D^0)$. The first decay is due to the quark transition 
$\bar b\to\bar c u\bar s$ proportional to $V_{cb}^* V_{us}$, whereas 
the second one is due to the decay $\bar b\to\bar u c\bar s$ 
proportional to $V_{ub}^* V_{cs}$. The relative phase of these two 
combinations of CKM elements is $\gamma$. Ideally, one would measure 
the six rates for the decays $B^+\to\bar D^0 K^+$, $B^+\to D^0 K^+$, 
$B^+\to D_1^0 K^+$, and their CP conjugates. Then, using isospin 
triangles, $\gamma$ could be determined in a theoretically clean way. 

This strategy is hindered by the fact that is it not possible 
experimentally to determine the rate of the doubly Cabibbo-suppressed 
decay $B^+\to D^0 K^+$ followed by $D^0\to K^-\pi^+$, because its 
combined branching ratio is similar to that of the transition 
$B^+\to\bar D^0 K^+$ followed by the doubly Cabibbo-suppressed decay 
$\bar D^0\to K^-\pi^+$ \cite{ADS}. Several approaches have been 
suggested to circumvent this problem \cite{ADS,Sinh}; however, they 
are challenging from an experimental point of view because precise 
measurements of very small decay rates are required.

Recently, some authors have suggested to use isospin relations
combined with certain dynamical assumptions (the neglect of
annihilation contributions relative to color-suppressed tree 
amplitudes) to eliminate the ``difficult'' $B^+\to D^0 K^+$ and 
$B^-\to\bar D^0 K^-$ rates in favor of six other $B,\bar B\to D K$ 
rates \cite{JaKo,GrJon}. Unfortunately, it is difficult to gauge 
the accuracy of the assumptions made in these proposals.

Considering the various options that have been discussed it appears 
that measuring $\gamma$ at the first-generation $B$ factories 
using $B\to D K^{(*)}$ decays is very challenging \cite{Soff}, and 
more demanding than a determination based on $B^\pm\to\pi K$ decays. 
On the other hand, to have two {\em independent\/} determinations 
of $\gamma$ from these two classes of decays would be extremely 
important. Whereas $\gamma$ measured in $B\to D\,K^{(*)}$ decays is 
likely to be the ``true'' phase of the CKM matrix, the angle 
$\gamma_{\pi K}$ determined in $B^\pm\to\pi K$ decays probes loop 
processes and may easily be affected by New Physics. As discussed 
in Section~\ref{sec:NP} and summarized in Table~\ref{tab:1}, 
comparing the two determinations would provide a very sensitive 
probe for physics beyond the Standard Model.

\section{Conclusions}

Among the strategies for determining the weak phase $\gamma
=\mbox{arg}(V_{ub}^*)$ of the quark mixing matrix, approaches based 
on rate measurements in $B\to\pi K$ decays play an important role 
and have received a lot of attention recently. The corresponding 
branching ratios are ``large'' and the final states are ``easy'' to 
detect experimentally. Using isospin, Fierz and flavor symmetries 
together with the fact that nonleptonic $B$ decays into two light 
mesons admit a heavy-quark expansion, a largely model-independent 
description of certain observables concerning the charged modes 
$B^\pm\to\pi K$ can be obtained. Various proposals exist for extracting 
information on $\gamma$ and on the Wolfenstein parameters $\bar\rho$ 
and $\bar\eta$ using these decays. In the future, this will allow us 
to determine $\gamma$ with a theoretical uncertainty of about 
$10^\circ$. When combined with an alternative measurement of $\gamma$ 
using other decays such as $B\to D K^{(*)}$, this will provide for 
a sensitive probe of physics beyond the Standard Model.

\acknowledgments
It is a pleasure to thank the SLAC Theory Group for the hospitality 
extended to me during my visit earlier this year. I am grateful 
to M. Beneke, G. Buchalla, Y. Grossman, A. Kagan, J. Rosner and C. 
Sachrajda for collaboration on parts of the work reported here. This 
research was supported by the Department of Energy under contract 
DE--AC03--76SF00515.

\end{document}